\documentclass[traditabstract,letter]{aa}

\usepackage{amsmath}
\usepackage{graphicx}
\usepackage{txfonts}
\usepackage{mathrsfs}
\usepackage{natbib}
\usepackage{color}
\usepackage{soul}
\usepackage{footmisc}
\usepackage{url}

\newcommand{\argmax}[1]{\underset{#1}{\operatorname{argmax}}}
\urldef\simbadhome\url{http://simbad.u-strasbg.fr/simbad/}
\urldef\twomasshome\url{http://www.ipac.caltech.edu/2mass/}

\begin{document}

\title{The radius and mass of the close solar twin 18 Sco derived from
    asteroseismology and interferometry\thanks{Based on
observations collected at the European Southern Observatory (ID
183.D-0729(A)) and at the CHARA Array, operated by Georgia State
University.}}

\author{M.~Bazot\inst{1}
\and M.~J.~Ireland\inst{2}
\and D.~Huber\inst{2}
\and T.~R.~Bedding\inst{2}
\and A.-M.~Broomhall\inst{3}
\and T.~L.~Campante\inst{1,4,5}
\and H.~Carfantan\inst{6}
\and W.~J.~Chaplin\inst{3}
\and Y.~Elsworth\inst{3}
\and J.~Mel\'endez\inst{1,7}
\and P.~Petit\inst{6}
\and S.~Th\'eado\inst{6}
\and V.~Van Grootel\inst{6}
\and T.~Arentoft\inst{4}
\and M.~Asplund\inst{8}
\and M.~Castro\inst{9}
\and J.~Christensen-Dalsgaard\inst{4}
\and J.~D.~do Nascimento Jr\inst{9}
\and B.~Dintrans\inst{6}
\and X.~Dumusque\inst{1,10}
\and H.~Kjeldsen\inst{4}
\and H. A. McAlister\inst{11}
\and T.~S.~Metcalfe\inst{12}
\and M.~J.~P.~F.~G.Monteiro\inst{1,5}
\and N.~C.~Santos\inst{1,5}
\and S.~Sousa\inst{1}
\and J.~Sturmann\inst{11}  
\and L.~Sturmann\inst{11} 
\and T.~A.~ten Brummelaar\inst{11}
\and N.~Turner\inst{11} 
\and S.~Vauclair\inst{6}
}

\institute{Centro de Astrof\'{\i}sica da Universidade do Porto, Rua das
  Estrelas, 4150-762, Porto, Portugal; bazot@astro.up.pt 
\and Sydney Institute for Astronomy (SIfA), School of Physics, University of Sydney NSW 2006, Australia 
\and School of Physics and Astronomy, University of Birmingham, Edgbaston, Birmingham B15 2TT, United-Kingdom 
\and Institut for Fysik og Astronomi, Aarhus Universitet, Ny Munkegade 1520, DK-8000 Aarhus C, Danmark 
\and Departamento de F\'{\i}sica e Astronomia, Faculdade de Ci\^encias, Universidade do Porto, Portugal
\and Laboratoire Astrophysique de Toulouse - Tarbes, Universit\'e de Toulouse, CNRS, Toulouse, France 
\and Departamento de Astronomia do IAG/USP, Universidade de S\~ao Paulo, Rua do Mat\~ao 1226, S\~ao Paulo, 05508-900, SP, Brasil 
\and Max Planck Institute for Astrophysics, Karl-Schwarzschild-Str. 1, Postfach 1317 D-85741 Garching, Germany 
\and Universidade Federal do Rio Grande do Norte, Dept de F\'{\i}sica Te\'orica e Experimental, Natal, 59072-970, RN, Brasil
\and Observatoire de Gen\`eve, 51 Chemin des Maillettes, CH-1290, Sauverny, Suisse
\and Center for High Angular Resolution Astronomy, Georgia State University, PO Box 3965, Atlanta, Georgia 30302-3965, USA
\and High Altitude Observatory, NCAR, Boulder, CO 80307, United States } 

\abstract{The growing interest in solar twins is motivated by
  the possibility of comparing them directly to the Sun. To carry on this
  kind of analysis, we need to know their physical
  characteristics with precision. Our first objective is to use
  asteroseismology and interferometry on the brightest of them: 18 Sco. We
  observed the star during 12 nights with HARPS for seismology and used the
  PAVO beam-combiner at CHARA for interferometry. An average large frequency separation $134.4\pm0.3$~$\mu$Hz and angular and linear radiuses of $0.6759 \pm 0.0062$~mas and $1.010\pm0.009$~R$_{\odot}$ were estimated. We used these values to derive the mass of the star, $1.02\pm0.03$~M$_{\odot}$.}

\maketitle

\section{Introduction}

Solar twins, defined as spectroscopically identical to the Sun
\citep{CdS81}, are important because they allow precise
differential analysis relative to the Sun
\citep[]{Ramirez09,Melendez09}.  The brightest solar twin
is 18 Sco (HD~146233, HIP~79672; $V=5.5$), 
whose mean atmospheric parameters are
$T_{\mathrm{eff}} = 5813 \pm 21$~K, $\log g = 4.45 \pm 0.02$ and
$\mathrm{[Fe/H]} = 0.04 \pm 0.01$ \citep{Takeda09,Ramirez09,Sousa08,Melendez07,Takeda07,Melendez06,Valenti05}.
  Its rotation rate
and magnetic field are also similar to solar ones \citep{Petit08}.  Its
position in the H-R diagram indicates that the star should be slightly younger and
more massive than the Sun \citep[][and references therein]{doNascimento09}.

During the past decade, asteroseismology and interferometry have arisen as
powerful techniques for constraining stellar parameters
\citep[e.g.,][]{Cunha07,Creevey07}.  Asteroseismology involves measuring the global
oscillation modes of a star (which for Sun-like stars are pressure-driven p
modes).  It is the only observational technique that is
directly sensitive to the deeper layers of the stellar interior, since the
characteristics of the modes depend on the regions through which the waves
travel.  Interferometry requires long-baseline interferometers capable of
resolving distant stars, hence allowing measurement of their
radii.

These techniques have already been combined to study the bright sub-giant
$\beta$~Hyi \citep{North07}, for which a mass was derived through homology
relations.  Here, we apply a similar method to 18 Sco.  In
Section~\ref{astero} we present the asteroseismic data and describe the
method used to derive the average large frequency separation.  In
Section~\ref{interfero} we describe the interferometric measurements that,
combined with the parallax, allow us to estimate the radius.  In
Section~\ref{mass} we use these quantities to derive the mass.

\section{Asteroseismology}\label{astero}

\begin{figure}[t!]
\center
\includegraphics[width=\columnwidth]{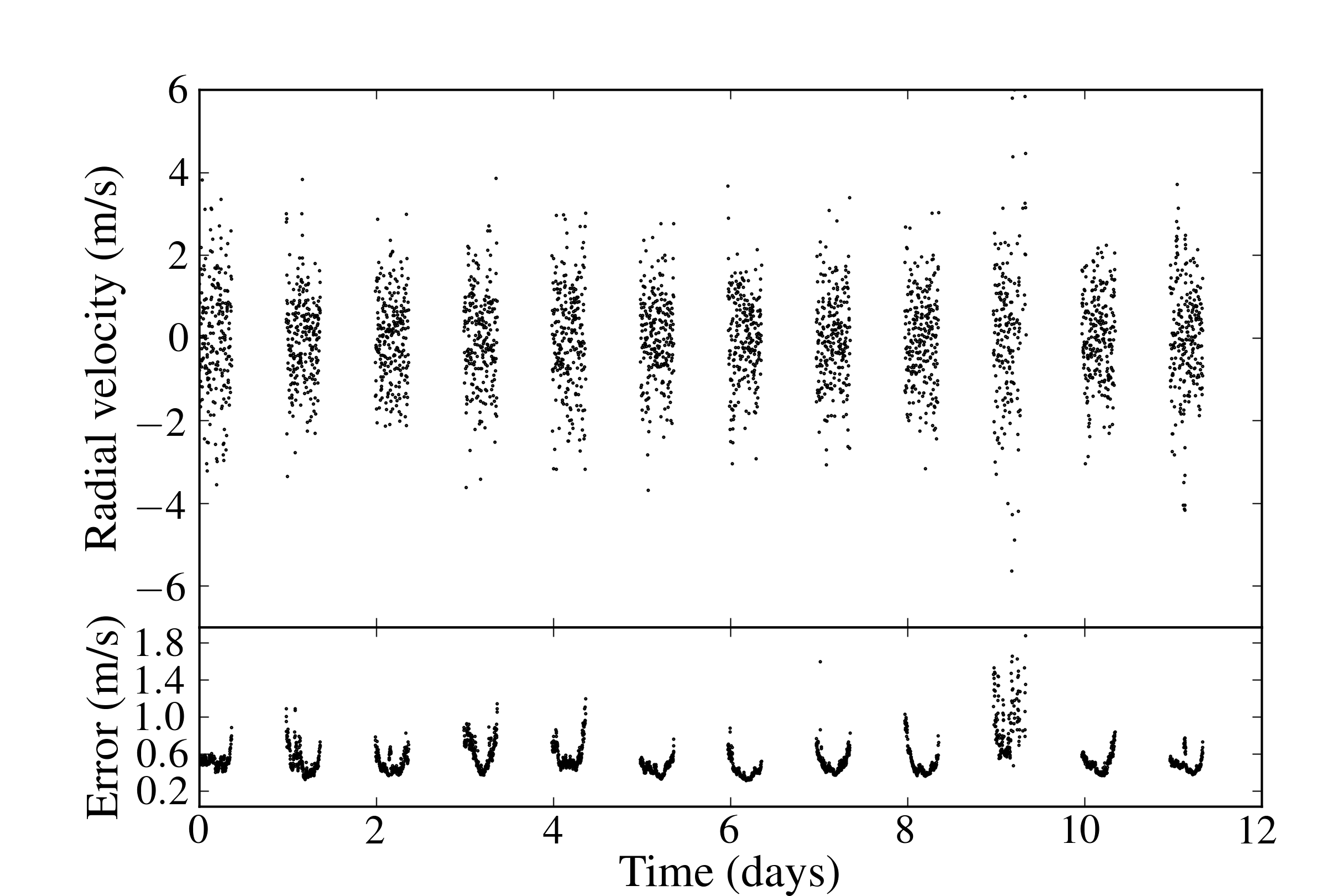}
\caption{Time series of radial velocities (upper panel) and their
  uncertainties (lower panel) from HARPS observations of 18 Sco.}
\label{rv}
\end{figure}

Detecting solar-like oscillations in a fifth-magnitude star from the ground is challenging and only a few instruments offer the required efficiency and high precision. We observed 18 Sco using the HARPS spectrograph on the 3.6-m telescope at
La Silla Observatory, Chile \citep{Mayor03}.  The data were collected over 12 nights from
10 to 21 May 2009\footnote{An attempt to reduce the daily aliases, with simultaneous observations on SOPHIE (Observatoire de Haute-Provence, run ID: 09A.PNPS.THEA), was unfortunately plagued by bad weather.}.
 We used the
high-efficiency mode with an average exposure time of 99.6~s. This resulted
in a typical signal-to-noise ratio at 550~nm of 158, with some exposures
reaching as high as~240.
The measured radial-velocity time series (filtered for low-frequency variations) is shown in Fig.~\ref{rv}.  We
obtained 2833 points with uncertainties in general below 2~m~s$^{-1}$. 
The average dispersion per night is $\sim$1.11~m~s$^{-1}$ and can be attributed mostly to p modes. 

\begin{figure}[t!]
\center
\includegraphics[width=\columnwidth]{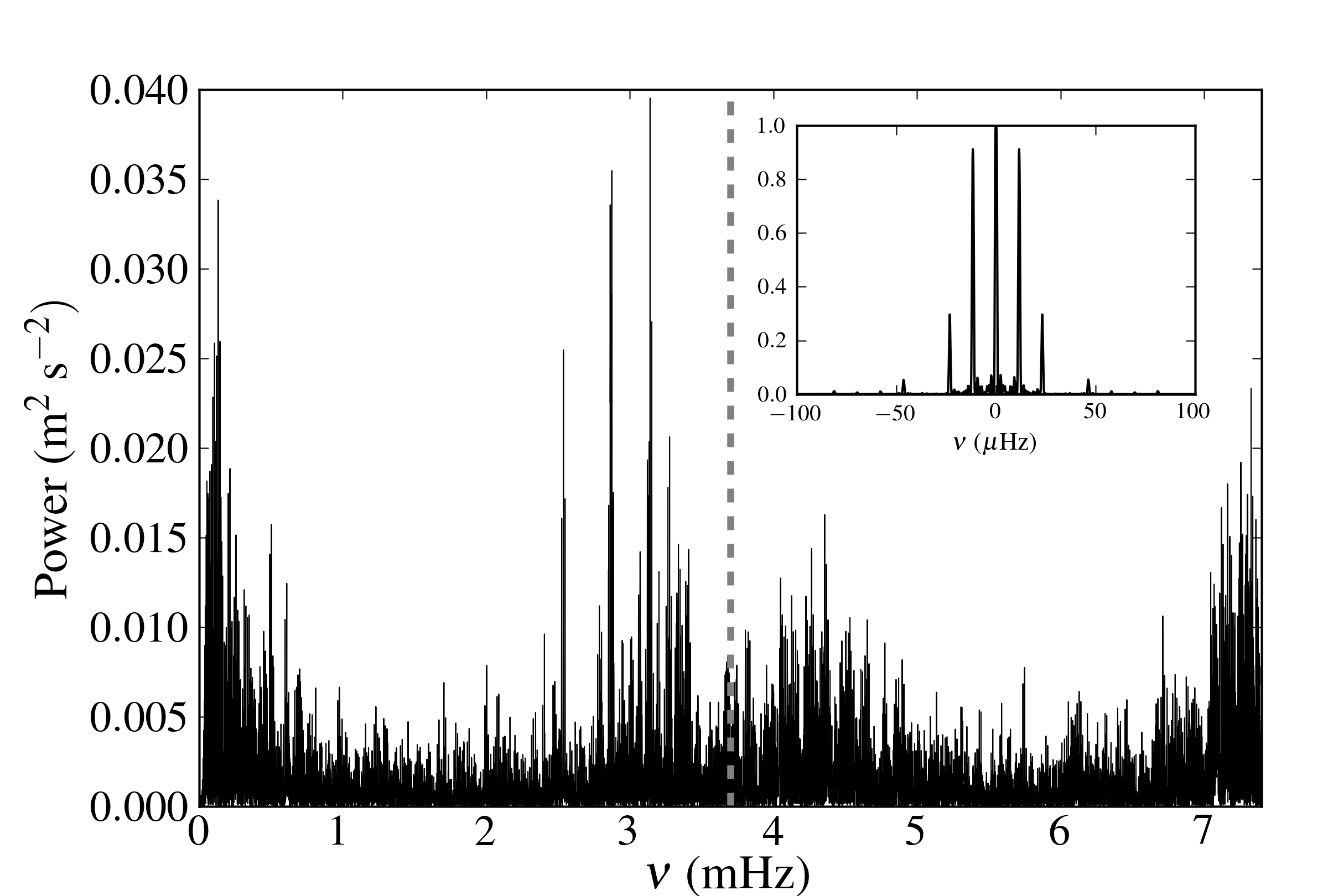}
\caption{Power spectrum of 18 Sco, evaluated using a weighted
Lomb-Scargle {\textquotedblleft}periodogram{\textquotedblright}. The
vertical grey dashed line marks the location of the equivalent Nyquist
frequency.  The inset shows the spectral window $W$, normalized to its maximum. }
\label{spectrum}
\end{figure}

The power spectrum, calculated using the measurement uncertainties as weights, is shown in Fig.~\ref{spectrum}.
The spectral window $W$, which is the Fourier transform of the observing
window,  $w(t)=\sum\delta(t-t_i) $ (with $t_i$ the mid-exposure time of
  the $i$-th exposure), is shown in the inset of Fig.~\ref{spectrum}.  Strong aliases caused by the daily gaps appear on both sides of the central peak at
multiples of $\pm 11.57$~$\mu$Hz.  The power spectrum shows a clear
excess around 3~mHz that is characteristic of solar-like oscillations,
reaching $\sim$0.04~m$^2$~s$^{-2}$ (corresponding to amplitudes $\sim$20~cm~s$^{-1}$).

The median sampling time was 135.0~s, including the read-out
($\sim$~22.6~s). This leads to an equivalent Nyquist
frequency of 3.7~mHz.  In Fig.~\ref{spectrum}, we clearly see a steep
rise in power at 7~mHz, corresponding to the folded low-frequency
increase. This also causes the bump that appears between 3.7~mHz and $\sim$5.5~mHz, which is an alias of the oscillation spectrum.

\begin{figure}
\center
\includegraphics[width=\columnwidth]{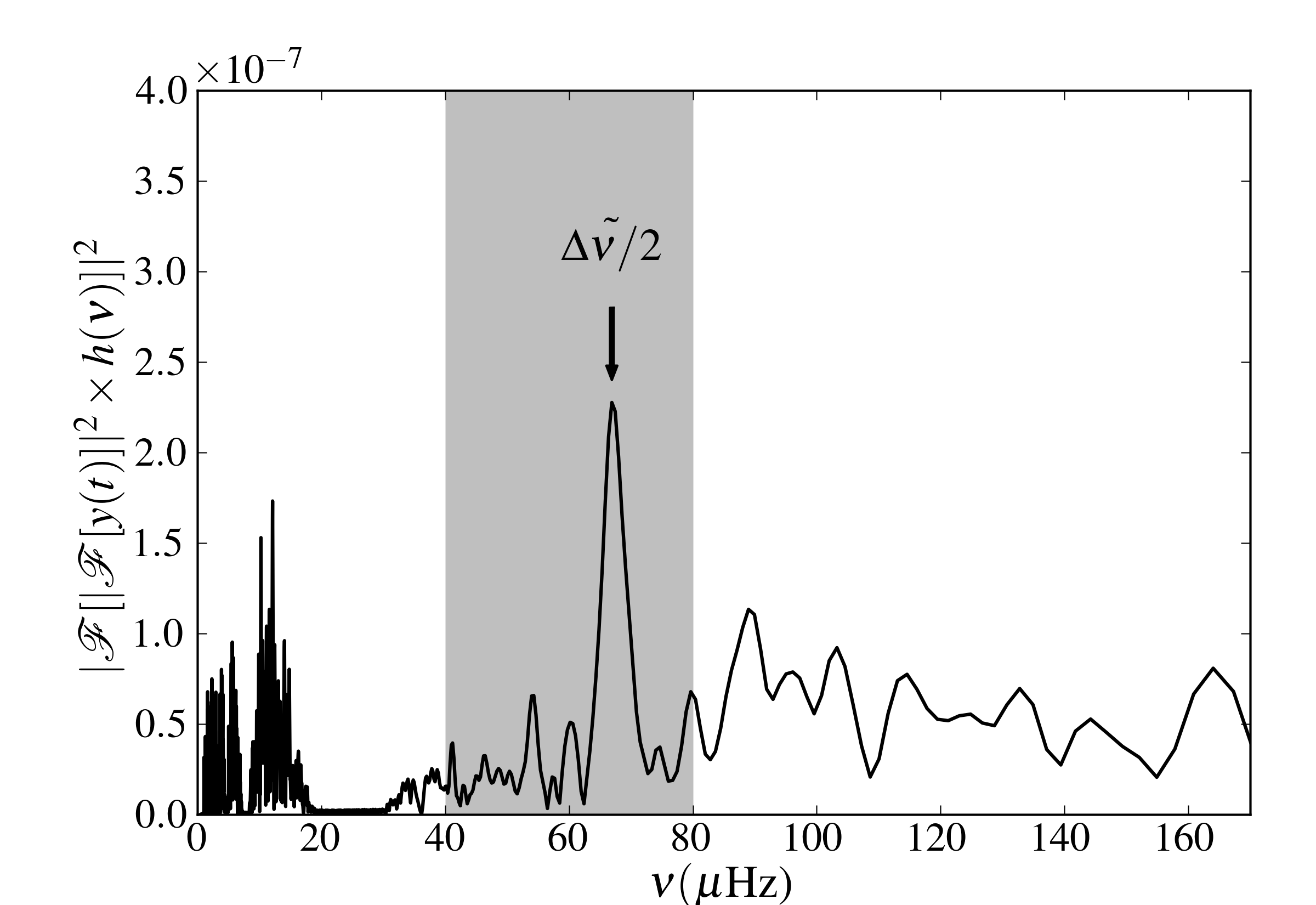}
\caption{Filtered autocorrelation function for the observed data. The shaded area marks the interval $\mathscr{T}$, in which we searched for the local maximum corresponding to $\Delta\nu/2$. $\mathscr{F}$ is the Fourier transform.}
\label{ACF1}
\end{figure}

The asymptotic relation for high-order p modes is $\nu_{n,l} =
(n+l/2+\epsilon_s)\Delta\nu$, where $\nu_{n,l}$ is the frequency of the mode
with radial order $n$ and angular degree $l$, the average large separation $\Delta\nu$ and a surface phase offset $\epsilon_s$ \citep{Tassoul80}. 
There is a periodicity of $\Delta\nu/2$ in the frequency
distribution, implying that there will be a local maximum at this value in the
autocorrelation function (ACF) of the signal $y(t)$, defined by
$R_{yy}(\tau) = E[y^{\ast}(t)y(t+\tau)]$ (where $E$ is the expectation
value and $y^{\ast}$ the complex conjugate of $y$). 
The ACF for 18 Sco is shown in
Fig.~\ref{ACF1}.  Since the signal is irregularly sampled with daily
gaps, it was computed by applying the Wiener-Khinchine theorem.

The large separation estimator is thus simply
\begin{equation}\label{estACF}
\tilde{\Delta \nu} = \displaystyle 2\times[\argmax{t\in\mathscr{T}} (R_{yy}\ast H)]^{-1},
\end{equation}
where $\mathscr{T}$ is the domain in which we search for this maximum,
which we set at $\mathscr{T}=[13000,25000]$~s (i.e. in the frequency range
40-80~$\mu$Hz), and $H$ the Fourier transform of a filter $h(\nu)$ that truncates the power spectrum.  Indeed, only the range 1500-3700~$\mu$Hz is considered when computing (using an FFT algorithm) the Fourier transform of the spectrum.  We used zero-padding to ensure that the ACF was evaluated at points separated by an {\textquotedblleft}equivalent frequency resolution{\textquotedblright} $\sim$0.01~$\mu$Hz.
 As noted by
\citet{Roxburgh09}, the width of $h(\nu)$ affects the localization of
$\tilde{\Delta \nu}$, which is a limitation of the method. 

The next step is to obtain information about the statistical properties of
the estimator $\tilde{\Delta \nu}$ that accounts for the noise in the data.
We can write
\begin{equation}
\pmb{y}=\pmb{x}+\pmb{\varepsilon},
\end{equation}
where $\pmb{y}=[y_0,\dots,y_N]$ are the measured values of the radial
velocity at times $t_0,\dots,t_N$, $\pmb{x}=[x_{0}\dots,x_N]$ are the true
values of the radial velocity and $\pmb{\varepsilon}=[\varepsilon_0,\dots,\varepsilon_N]$ is a vector
gathering the noise contributions from observational
errors.  Our goal is to estimate the probability
density of $\tilde{\Delta \nu}$ conditional on $\pmb{y}$, $p(\tilde{\Delta
\nu} | \pmb{y})$. To do so, we used a Monte Carlo approach to error
propagation. We assumed that the noise in the data is a series of
realizations of independent random variables distributed according to the
Gaussian distributions $\mathcal{N}_i=\mathcal{N}(x(t_i),\sigma_i^2)$, with
the $\sigma_i$ given by the uncertainties on the data.  We
simulated time series by generating new realizations of the
noise distributed according to the $\mathcal{N}_i$ at each $t_i$ and adding them to
$y_i$. For each artificial set of data $\pmb{y^a}$, we estimated
$\tilde{\Delta\nu^a}$.

Our process for generating the artificial data means that it satisfies
\begin{equation}
\pmb{y^a}=\pmb{y}+\pmb{\varepsilon'}=\pmb{x}+\pmb{\varepsilon}+\pmb{\varepsilon'},
\end{equation}
with $\pmb{\varepsilon'}$ the artificially generated noise, and $\pmb{\varepsilon'}$ and $\pmb{\varepsilon}$ both realizations of the same distribution at time
$t_i$. Ideally, one wishes to estimate the large separation from $\pmb{x}$. One
possibility would be to estimate it from several measurements, on
different telescopes at the same times $t_i$. A second way would be
to generate the $\pmb{y^a}$ from a model reproducing the data
$\pmb{y}$, then perturbing the output of this model, rather than the
real observations, which is the classical procedure of Monte Carlo estimation
of parameters. Unfortunately, the knowledge of $\Delta\nu$ alone does
not permit such a model to be set up. 

\begin{figure}[t!]
\center
\includegraphics[width=\columnwidth]{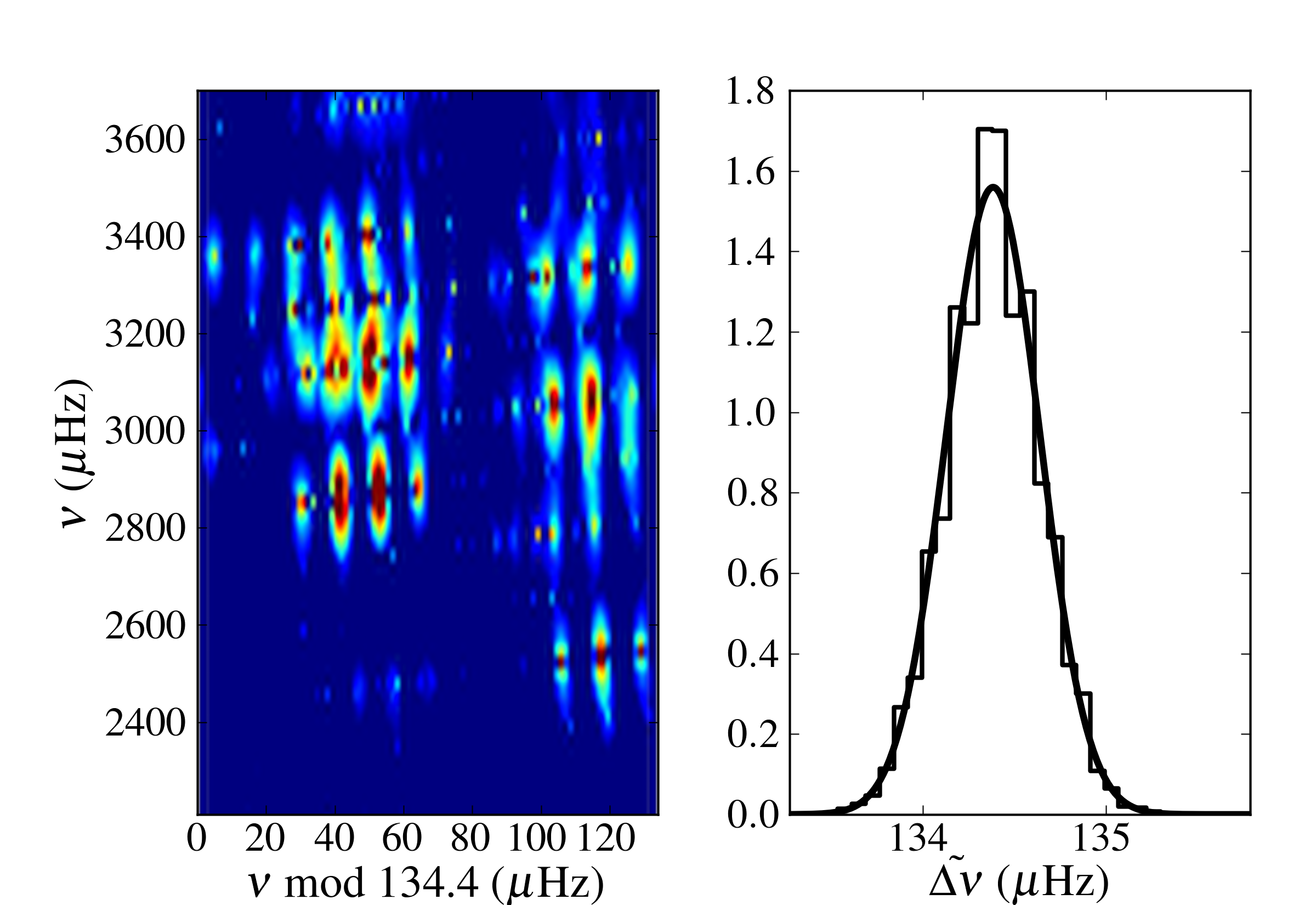}
\caption{ Left panel: \'echelle
diagram corresponding to the mean value of $\tilde{\Delta\nu}$. Right panel: results from the Monte Carlo experiment. The histogram shows the distribution of the actual realizations, and the continuous line the Gaussian with the corresponding mean and variance.}
\label{ACF2}
\end{figure}

 It thus has to be assumed that this
bias will not be too severe, which can be crudely checked graphically with an
\'echelle diagram (see Fig.~\ref{ACF2}). Further estimations of the large
separations, using individual frequencies, may give us some
information on its magnitude. However, the error bars are representative of the error propagation: were we able to correct for the bias induced by random observational
noise, we would expect our confidence interval on the large separation
to be the same.

Figure~\ref{ACF2} shows the results for our Monte Carlo suite of
10000 time series. It closely follows a Gaussian
distribution with parameters $\mu = 134.4$~$\mu$Hz and $\sigma =
0.3$~$\mu$Hz. The right panel represents the
\'echelle diagram of the observations using this value for the mean large
separation.

\section{Interferometry}\label{interfero}

To measure the angular diameter of 18 Sco, which is expected to only be 
about 0.7\,mas, we used long-baseline interferometry at visible
wavelengths.  We used the PAVO beam combiner (Precision Astronomical Visible
Observations; \citealt{Ireland08}) at the CHARA array (Center for High
Angular Resolution Astronomy; \citealt{tenBrummelaar05}).  We obtained four
calibrated sets of observations on 18 July 2009 using the S1-W2 (211~m)
baseline.

PAVO records fringes in 38 wavelength channels centred on the $R$ band
($\lambda_c \simeq 700$~nm).  The raw data were reduced using the PAVO data
analysis pipeline (M. J. Ireland et al. in preparation).  To enhance the signal-to-noise
ratio, the analysis pipeline can average over several wavelength channels,
and for 18 Sco we found an optimal smoothing width of five channels.
Excluding four channels on each end because of edge effects, this resulted in
six independent data points per scan and hence a total of 24 independent visibility
measurements for 18 Sco.

\begin{figure}
\center
\includegraphics[width=\columnwidth]{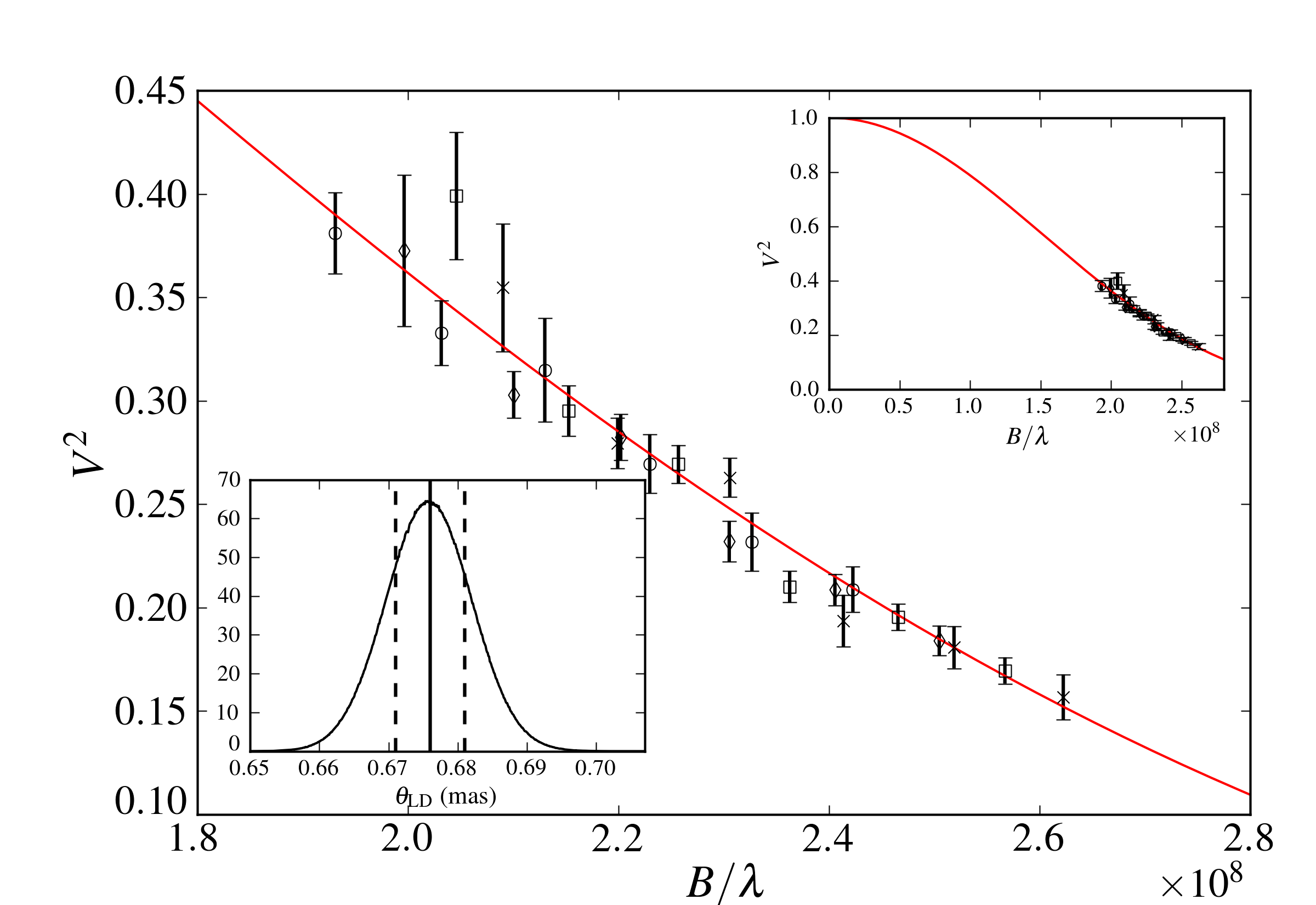}
\sethlcolor{red}
\caption{Calibrated squared visibilities for 18 Sco. The red line represents
the best model. Each symbol type corresponds to one scan, the visibility averaged over five wavelength channels. Upper-right inset: Same model represented on a larger scale. Lower-left inset: distribution for angular diameter (mean and standard deviation represented by vertical lines).  }
\label{interf}
\sethlcolor{yellow}
\end{figure}

Table~\ref{calib} lists the three stars we used to calibrate the
visibilities.  We estimated their angular diameters, $\theta$, using the $V$--$K$
calibration of \citet{Kervella04}.  Although the internal precision of this calibration, as well as the uncertainties in the photometry for all three stars, is better than 1\%, we assume here conservative uncertainties of 5\% for each calibrator \citep{vanBelle05}. These incorporate the unknown orientation and expected oblateness in fast rotators \citep{Royer02}. These were chosen to be
single stars with predicted diameters at least a factor of two smaller than
18 Sco and to be nearby on the sky (at a separation $d<10\degr$).  Each of the four scans
of 18 Sco was calibrated, using a weighted mean of the calibrators
bracketing the scan.  All scans contributing to a bracket were made within
a time interval of 15 minutes.  The final calibrated squared-visibility
measurements are shown in Fig.~\ref{interf} as a function of
spatial frequency.

To determine the angular diameter, corrected for limb darkening, we fitted
the following model to the data \citep{HanburyBrown74}:
\begin{equation}
V= \left( \frac{1-\mu_{\lambda}}{2}+\frac{\mu_{\lambda}}{3}\right)^{-1} \left[
(1-\mu_{\lambda})\frac{J_1(x)}{x} + \mu_{\lambda} (\pi/2)^{1/2}\frac{J_{3/2}(x)}{x^{3/2}}\right]
\end{equation}  
with $x=\pi B\theta_{\mathrm{LD}}\lambda^{-1}$. Here, $V$ is the
visibility, $\mu_{\lambda}$ the linear limb-darkening coefficient,
$J_n(x)$ the $n$-th order Bessel function, $B$ the projected
baseline, $\theta_{\mathrm{LD}}$ the limb-darkened angular diameter, and
$\lambda$ the wavelength at which the observations were done. In our
analysis, we used $\mu_{\lambda} = 0.607 \pm 0.012$, which is interpolated at the $T_{\mathrm{eff}}$, $\log g$,
and metallicity of 18 Sco in the R-filter given in the catalog of
\citet{Claret00}. For all wavelength channels, we assumed an absolute error of 5nm ($\sim$0.5\%).

Interferometric measurements are often dominated by systematic errors and therefore require a careful analysis of all error sources. To arrive at realistic uncertainties for the angular diameter, we performed a series of $10^4$ Monte Carlo simulations as follow. For each simulation, we drew realizations from the observed values (assuming they correspond  to the parameters of Gaussian distributions) for the calibrator angular diameters, limb darkening coefficient, and wavelength channels. With these parameters we then calibrated the raw visibility measurements and fit the angular diameter $\theta_{\mathrm{LD}}$ to the calibrated data using a least-squares minimization algorithm. Finally, we generated for each simulation a random sample of 200 normally distributed points with a mean corresponding to the fitted diameter and a standard deviation corresponding to the formal uncertainty (scaled so that $\chi^2=1$) of the fit. For each Monte-Carlo simulation these 200 points were stored to make up the final distribution containing $2\times10^6$ points. This procedure was carried out for all independent measurements in our data.

The resulting distribution for the diameter of 18 Sco is shown in Fig.~5, along with the best-fitting model. The mean and standard deviation of this distribution yield $\theta_{\mathrm{LD}} = 0.6759 \pm 0.0062$~mas. Combined with the Hipparcos parallax of $71.94\pm0.37$~mas \citep{vanLeeuwen07}, we find the radius of 18 Sco to be $R/R_{\odot} = 1.010\pm0.009$.  We conclude that the radius of 18 Sco is the same as the Sun, within an uncertainty of 0.9\%.

\begin{table}[t]
\begin{minipage}[h!]{\columnwidth}
\caption{Properties of the calibrators used for 18 Sco.}
\label{calib}      
\centering                      
\renewcommand{\footnoterule}{}
\begin{tabular}{lccccc}          
\hline\hline                       
Star & $V$\footnote{\simbadhome} & $K$ \footnote{\twomasshome}& Sp. Type&  $\theta$ (mas)& $d$ (deg)\\    
\hline                                   
HD145607& 5.435 & 5.052& A4V & 0.343$\pm$0.017 & 0.9\\      
HD145788& 6.255 & 5.737& A1V & 0.256$\pm$0.013 & 4.2\\
HD147550& 6.245 & 5.957& B9V & 0.233$\pm$0.012 & 6.5\\
\hline                                             
\end{tabular}
\end{minipage}	       
\end{table}

\section{Mass of 18 Sco}\label{mass}

\citet{Gough90} pointed out that the homology relation:
\begin{equation}\label{homology}
\Delta\nu \propto M^{1/2}R^{-3/2},
\end{equation}
 holds for main-sequence stars even outside the zero-age main sequence. Other studies have confirmed this picture \citep[e.g.,][]{Stello09}.  

Applying the ACF method described above to a one-year BiSON
time series of the Sun \citep[][and references therein]{Broomhall09}, we found the solar large separation to be
$135.229\pm0.003$~$\mu$Hz.  
With our radius measurement this gives a mass $M=1.02\pm0.03$~M$_{\odot}$ for 18 Sco.  The agreement is good with the
published estimates derived from indirect methods, such as comparison between
spectro- or photometric observations and stellar evolutionary tracks
\citep{Valenti05,Melendez07,Takeda07,Sousa08,doNascimento09}.

The assumption of homology is in general well-supported by the comparison
to models. Considering a small departure from it in the form $\Delta\nu\propto(1+c)M^{1/2}R^{-3/2}$ (with $ c \ll
1$, being a function of the structure of the star), we then have for a
scaling relative to the Sun,
$c=(\Delta\nu/\Delta\nu_{\odot})(M/M_{\odot})^{1/2}(R/R_{\odot})^{-3/2}-1$. For
stellar ages characteristics of those quoted for 18 Sco (the dependence on
the mass and the metallicity being weak), this quantity may contribute to
an additional $\sim$0.2\%-0.4\% on the total error on the mass.

The impact of filtering the ACF is not completely negligible, and if we vary
the lower limit of $\mathscr{T}$, between 1500~$\mu$Hz and 2000~$\mu$Hz,
the final estimate may vary by $\sim$0.01~M$_{\odot}$, emphasizing the need
for individual frequency measurements.

\section{Conclusion}

We presented the first asteroseismic and interferometric measurements for
the solar twin 18 Sco. These allowed us to estimate a mass for this star
independent of the previous spectro-photometric studies, which are still being
confirmed. This work shows the possibilities offered by asteroseismology,
even from a ground-based single site, and iby nterferometry. Our results confirm that 18 Sco is remarkably similar to the Sun in both radius and mass.

The next step will involve measuring the individual oscillation frequencies
and performing full modelling using all the available observations. It will hopefully reduce the uncertainty on the estimated age, improving our knowledge of the physical state of 18 Sco \citep{doNascimento09}. This will provide a more precise picture of its interior and give information on the depth of its external convective zone \citep{Monteiro00}, which is
necessary if one wishes to study its magnetic activity cycle
\citep{Petit08}.

\begin{acknowledgements}
This work was supported by grants SFRH/BPD/47994/2008, PTDC/CTE-AST/098754/2008, and PTDC/CTE-AST/66181/2006, from FCT/MCTES and FEDER, Portugal. This research was supported by the Australian Research Council (project number DP0878674). Access to CHARA was funded by the AMRFP (grant 09/10-O-02), supported by the Commonwealth of Australia under the International Science Linkages programme. 
The CHARA Array is owned by Georgia State University. Additional funding for the CHARA Array is provided by the National Science Foundation under grant AST09-08253, by the W. M. Keck Foundation, and the NASA Exoplanet Science Center.
\end{acknowledgements}

\bibliography{18scoref}

\begin{thebibliography}{29}
\expandafter\ifx\csname natexlab\endcsname\relax\def\natexlab#1{#1}\fi

\bibitem[{{Broomhall} {et~al.}(2009){Broomhall}, {Chaplin}, {Davies},
  {Elsworth}, {Fletcher}, {Hale}, {Miller}, \& {New}}]{Broomhall09}
{Broomhall}, A., {Chaplin}, W.~J., {Davies}, G.~R., {et~al.} 2009, \mnras, 396,
  L100

\bibitem[{{Cayrel de Strobel} {et~al.}(1981){Cayrel de Strobel}, {Knowles},
  {Hernandez}, \& {Bentolila}}]{CdS81}
{Cayrel de Strobel}, G., {Knowles}, N., {Hernandez}, G., \& {Bentolila}, C.
  1981, \aap, 94, 1

\bibitem[{{Claret}(2000)}]{Claret00}
{Claret}, A. 2000, \aap, 363, 1081

\bibitem[{{Creevey} {et~al.}(2007){Creevey}, {Monteiro}, {Metcalfe}, {Brown},
  {Jim{\'e}nez-Reyes}, \& {Belmonte}}]{Creevey07}
{Creevey}, O.~L., {Monteiro}, M.~J.~P.~F.~G., {Metcalfe}, T.~S., {et~al.} 2007,
  \apj, 659, 616

\bibitem[{{Cunha} {et~al.}(2007){Cunha}, {Aerts}, {Christensen-Dalsgaard},
  {Baglin}, {Bigot}, {Brown}, {Catala}, {Creevey}, {Domiciano de Souza},
  {Eggenberger}, {Garcia}, {Grundahl}, {Kervella}, {Kurtz}, {Mathias},
  {Miglio}, {Monteiro}, {Perrin}, {Pijpers}, {Pourbaix}, {Quirrenbach},
  {Rousselet-Perraut}, {Teixeira}, {Th{\'e}venin}, \& {Thompson}}]{Cunha07}
{Cunha}, M.~S., {Aerts}, C., {Christensen-Dalsgaard}, J., {et~al.} 2007, \aapr,
  14, 217

\bibitem[{{do Nascimento} {et~al.}(2009){do Nascimento}, {Castro},
  {Mel{\'e}ndez}, {Bazot}, {Th{\'e}ado}, {Porto de Mello}, \& {de
  Medeiros}}]{doNascimento09}
{do Nascimento}, Jr., J.~D., {Castro}, M., {Mel{\'e}ndez}, J., {et~al.} 2009,
  \aap, 501, 687

\bibitem[{{Gough}(1990)}]{Gough90}
{Gough}, D.~O. 1990, in Astrophysics: Recent Progress and Future Possibilities,
  ed. {B.~Gustafsson \& P.~E.~Nissen}, 13--50

\bibitem[{{Hanbury Brown} {et~al.}(1974){Hanbury Brown}, {Davis}, {Lake}, \&
  {Thompson}}]{HanburyBrown74}
{Hanbury Brown}, R., {Davis}, J., {Lake}, R.~J.~W., \& {Thompson}, R.~J. 1974,
  \mnras, 167, 475

\bibitem[{{Ireland} {et~al.}(2008){Ireland}, {M{\'e}rand}, {ten Brummelaar},
  {Tuthill}, {Schaefer}, {Turner}, {Sturmann}, {Sturmann}, \&
  {McAlister}}]{Ireland08}
{Ireland}, M.~J., {M{\'e}rand}, A., {ten Brummelaar}, T.~A., {et~al.} 2008, in
  Optical and Infrared Interferometry, Proc.\ SPIE vol.\ 7013, 701324

\bibitem[{{Kervella} {et~al.}(2004){Kervella}, {Th{\'e}venin}, {Di Folco}, \&
  {S{\'e}gransan}}]{Kervella04}
{Kervella}, P., {Th{\'e}venin}, F., {Di Folco}, E., \& {S{\'e}gransan}, D.
  2004, \aap, 426, 297

\bibitem[{{Mayor} {et~al.}(2003){Mayor}, {Pepe}, {Queloz}, {Bouchy},
  {Rupprecht}, {Lo Curto}, {Avila}, {Benz}, {Bertaux}, {Bonfils}, {Dall},
  {Dekker}, {Delabre}, {Eckert}, {Fleury}, {Gilliotte}, {Gojak}, {Guzman},
  {Kohler}, {Lizon}, {Longinotti}, {Lovis}, {Megevand}, {Pasquini}, {Reyes},
  {Sivan}, {Sosnowska}, {Soto}, {Udry}, {van Kesteren}, {Weber}, \&
  {Weilenmann}}]{Mayor03}
{Mayor}, M., {Pepe}, F., {Queloz}, D., {et~al.} 2003, The Messenger, 114, 20

\bibitem[{{Mel{\'e}ndez} {et~al.}(2009){Mel{\'e}ndez}, {Asplund}, {Gustafsson},
  \& {Yong}}]{Melendez09}
{Mel{\'e}ndez}, J., {Asplund}, M., {Gustafsson}, B., \& {Yong}, D. 2009, \apjl,
  704, L66

\bibitem[{{Mel{\'e}ndez} {et~al.}(2006){Mel{\'e}ndez}, {Dodds-Eden}, \&
  {Robles}}]{Melendez06}
{Mel{\'e}ndez}, J., {Dodds-Eden}, K., \& {Robles}, J.~A. 2006, \apjl, 641, L133

\bibitem[{{Mel{\'e}ndez} \& {Ram{\'{\i}}rez}(2007)}]{Melendez07}
{Mel{\'e}ndez}, J. \& {Ram{\'{\i}}rez}, I. 2007, \apjl, 669, L89

\bibitem[{{Monteiro} {et~al.}(2000){Monteiro}, {Christensen-Dalsgaard}, \&
  {Thompson}}]{Monteiro00}
{Monteiro}, M.~J.~P.~F.~G., {Christensen-Dalsgaard}, J., \& {Thompson}, M.~J.
  2000, \mnras, 316, 165

\bibitem[{{North} {et~al.}(2007){North}, {Davis}, {Bedding}, {Ireland},
  {Jacob}, {O'Byrne}, {Owens}, {Robertson}, {Tango}, \& {Tuthill}}]{North07}
{North}, J.~R., {Davis}, J., {Bedding}, T.~R., {et~al.} 2007, \mnras, 380, L80

\bibitem[{{Petit} {et~al.}(2008){Petit}, {Dintrans}, {Solanki}, {Donati},
  {Auri{\`e}re}, {Ligni{\`e}res}, {Morin}, {Paletou}, {Ramirez}, {Catala}, \&
  {Fares}}]{Petit08}
{Petit}, P., {Dintrans}, B., {Solanki}, S.~K., {et~al.} 2008, \mnras, 388, 80

\bibitem[{{Ram{\'{\i}}rez} {et~al.}(2009){Ram{\'{\i}}rez}, {Mel{\'e}ndez}, \&
  {Asplund}}]{Ramirez09}
{Ram{\'{\i}}rez}, I., {Mel{\'e}ndez}, J., \& {Asplund}, M. 2009, \aap, 508, L17

\bibitem[{{Roxburgh}(2009)}]{Roxburgh09}
{Roxburgh}, I.~W. 2009, \aap, 506, 435

\bibitem[{{Royer} {et~al.}(2002){Royer}, {Grenier}, {Baylac}, {G{\'o}mez}, \&
  {Zorec}}]{Royer02}
{Royer}, F., {Grenier}, S., {Baylac}, M., {G{\'o}mez}, A.~E., \& {Zorec}, J.
  2002, \aap, 393, 897

\bibitem[{{Sousa} {et~al.}(2008){Sousa}, {Santos}, {Mayor}, {Udry},
  {Casagrande}, {Israelian}, {Pepe}, {Queloz}, \& {Monteiro}}]{Sousa08}
{Sousa}, S.~G., {Santos}, N.~C., {Mayor}, M., {et~al.} 2008, \aap, 487, 373

\bibitem[{{Stello} {et~al.}(2009){Stello}, {Chaplin}, {Basu}, {Elsworth}, \&
  {Bedding}}]{Stello09}
{Stello}, D., {Chaplin}, W.~J., {Basu}, S., {Elsworth}, Y., \& {Bedding}, T.~R.
  2009, \mnras, 400, L80

\bibitem[{{Takeda} {et~al.}(2007){Takeda}, {Kawanomoto}, {Honda}, {Ando}, \&
  {Sakurai}}]{Takeda07}
{Takeda}, Y., {Kawanomoto}, S., {Honda}, S., {Ando}, H., \& {Sakurai}, T. 2007,
  \aap, 468, 663

\bibitem[{{Takeda} \& {Tajitsu}(2009)}]{Takeda09}
{Takeda}, Y. \& {Tajitsu}, A. 2009, \pasj, 61, 471

\bibitem[{{Tassoul}(1980)}]{Tassoul80}
{Tassoul}, M. 1980, \apjs, 43, 469

\bibitem[{{ten Brummelaar} {et~al.}(2005){ten Brummelaar}, {McAlister},
  {Ridgway}, {Bagnuolo}, {Turner}, {Sturmann}, {Sturmann}, {Berger}, {Ogden},
  {Cadman}, {Hartkopf}, {Hopper}, \& {Shure}}]{tenBrummelaar05}
{ten Brummelaar}, T.~A., {McAlister}, H.~A., {Ridgway}, S.~T., {et~al.} 2005,
  \apj, 628, 453

\bibitem[{{Valenti} \& {Fischer}(2005)}]{Valenti05}
{Valenti}, J.~A. \& {Fischer}, D.~A. 2005, \apjs, 159, 141

\bibitem[{{van Belle} \& {van Belle}(2005)}]{vanBelle05}
{van Belle}, G.~T. \& {van Belle}, G. 2005, \pasp, 117, 1263

\bibitem[{{van Leeuwen}(2007)}]{vanLeeuwen07}
{van Leeuwen}, F., ed. 2007, Astrophysics and Space Science Library, Vol. 350,
  {Hipparcos, the New Reduction of the Raw Data}

\end{thebibliography}

\end{document}